\documentclass[12pt,english]{paper}
\usepackage{amssymb}
\usepackage[T1]{fontenc}
\usepackage[latin1]{inputenc}
\usepackage{graphicx}
\usepackage{cite}
\makeatletter

\newcommand{\bibstyle@aas}{\bibpunct{(}{)}{;}{a}{}{,}}

\makeatletter

\makeatletter

\usepackage{geometry}

\makeatletter

\usepackage{epsfig}

\makeatother

\usepackage{babel}
\makeatother
\begin{document}

\title{Detection of Ultra-High Energy neutrinos skimming the Earth due to decay of superheavy dark matter at JEM-EUSO}

\author{Ye Xu$^{1,2}$}

\maketitle

\begin{flushleft}
$^1$School of Information Science and Engineering,  Fujian University of Technology, Fuzhou 350118, China
\par
$^2$Research center for Microelectronics Technology, Fujian University of Technology, Fuzhou 350118, China
\par
e-mail address: xuy@fjut.edu.cn
\end{flushleft}

\begin{abstract}
The possibility of detecting Ultra-high energy (UHE from now on) neutrinos due to superheavy dark matter are considered by the neutrinos interaction with the nuclei in the air in the present paper. To reject other standard model particles, UHE neutrinos, from superheavy dark matter and astrophysical sources, skimming the Earth are detected at JEM-EUSO. Then the numbers of UHE neutrinos detected by JEM-EUSO are evaluated at different energies (1 EeV < E < 1 ZeV) in 5 years. If the energy thresholds are taken to be 100 EeV, the astrophysical neutrino contamination could be negligible in this detection. It is possible that UHE neutrinos from the decay of superheavy dark matter are detected at JEM-EUSO when $O(10^{27})s < \tau_{\phi}< O(10^{31})s$. For example, O($10^4$) UHE neutrino events could be detected by JEM-EUSO at 30 EeV in 5 years when $\tau_{\phi}=O(10^{27})$s. If so, it may be confirmed that there may exist superheavy dark matter in the Universe.
\end{abstract}


\section{Introduction}
It is indicated by the Planck data with measurements of the cosmic microwave background
 that $26.6\%$ of the overall energy density of
the Universe is non-baryonic dark matter\cite{Planck2015}. They are
distributed in a halo surrounding a galaxy. A dark matter halo of a galaxy
with a local density of 0.3 GeV/cm$^3$ is assumed and its relative
speed to the Sun is 230 km/s\cite{JP}. At present, one mainly searches for thermal
dark matter particles via direct and indirect detections\cite{CDMSII,CDEX,XENON1T,LUX,PANDAX,AMS-02,DAMPE,fermi}. Because of the very small cross sections of the interactions
between these dark matter particles and nucleus (maybe O(10$^{-47}$ cm$^2$))\cite{XENON1T,PANDAX}, so far one has not found dark matter yet.
\par
It is considered an assumption that there is a non-thermal dark sector generated by the early Universe with its bulk comprised of a very massive relic $\phi$ in the Universe\cite{KC87, CKR98, CKR99, CGIT, FKMY}. This superheavy dark matter may decays to the Standard Model (SM from now on) particles and its lifetime is much greater than the age of the Universe. This lead to a small but significant flux of UHE SM particles. In the present work, it is only focused on detection of the UHE neutrinos induced by the decay of superheavy dark matter $\phi$ ($\phi\to\nu\bar{\nu}$)\cite{LT,EIP,BLS,BKMTZ}. Although the fraction of these relativistic neutrinos is small in the Universe, their large interaction cross sections between them and other SM particles make it possible to find them. If so, it is confirmed that there may exist superheavy dark matter in the Universe. These UHE neutrinos, which pass through the Earth and air and interact with nuclei, can be detected by JEM-EUSO\cite{JEM-EUSO}, via fluorescent and Cherenkov photons due to the development of extensive air showers. To reject other SM particles, UHE neutrinos skimming the Earth are detected at JEM-EUSO. In my work, it is confirmed the possibility that UHE neutrinos, from the decay of superheavy dark matter, skimming the Earth could be detected at JEM-EUSO . In this detection, the main contamination is from the astrophysical diffuse neutrinos in the University.
\par
In what follows, the UHE neutrino event rates, including from the decay of superheavy dark matter and astrophysical sources, will be estimated at JEM-EUSO. And it is discussed the possibility of detection of UHE neutrinos induced by the decay of superheavy dark matter.
\section{UHE neutrinos flux from the decay of superheavy dark matter}
It is considered a scenario where a co-moving non-relativistic scalar species $\phi$, with mass $m_{\phi} \geq$ 1 PeV, in the Universe\cite{EIP,BLS,BGA}. This dark matter decay to SM particles with a very large lifetime. The lifetime for the decay of superheavy dark matter to SM particles is strongly constrained ($\tau \geq$ O($10^{26}-10^{29}$)s) by diffuse gamma and neutrino observations\cite{EIP,MB,RKP,KKK}. And $\tau_{\phi}$ is taken to be $10^{27}$s in the present work. Then the UHE neutrino flux from the Galaxy is obtained via the following equation\cite{BLS}:
\begin{center}
\begin{equation}
\psi_{\nu}=\int_{E_{min}}^{E_{max}}F\frac{dN_\nu}{dE_\nu}dE
\end{equation}
\end{center}
with
\par
\begin{center}
\begin{equation}
F=1.7\times10^{-12}\times\frac{10^{28}s}{\tau_{\phi}}\times\frac{1PeV}{m_{\phi}}cm^{-2}s^{-1}sr^{-1}.
\end{equation}
\end{center}
where E$_{\nu}$ and N$_{\nu}$ are the energy and number of UHE neutrino, respectively. $\displaystyle\frac{dN_{\nu}}{dE_{\nu}}=2\delta(E_{\nu}-\displaystyle\frac{m_{\phi}}{2})$.
\section{UHE neutrino interaction with nuclei}
For neutrinos at energies above 1 PeV, their interaction cross sections with nucleus are given by simple power-law forms\cite{BHM}:
\begin{center}
\begin{equation}
\sigma_{\nu N}(CC)=4.74\times10^{-35} cm^2 \left(\frac{E_{\nu}}{1 GeV}\right)^{0.251}
\end{equation}
\end{center}
\begin{center}
\begin{equation}
\sigma_{\nu N}(NC)=1.80\times10^{-35} cm^2 \left(\frac{E_{\nu}}{1 GeV}\right)^{0.256}
\end{equation}
\end{center}
where $E_{\nu}$ is the neutrino energy.
\par
The neutrino interaction length can be obtained by
\par
\begin{center}
\begin{equation}
L_{\nu}=\frac{1}{N_A\rho\sigma_{\nu N}}
\end{equation}
\end{center}
\par
where $N_A$ is the Avogadro constant, and $\rho$ is the density of matter, which neutrinos interact with, and $\rho$ is taken to be 3 g/cm$^3$ in the case of detection of neutrinos skimming the Earth.
\section{Evaluation of the numbers of UHE neutrinos detected by JEM-EUSO}
UHE neutrinos are produced by decay of superheavy dark matter ($\phi\to\nu\bar{\nu}$) and their energies depend on the mass of $\phi$ (that is $E_{\nu}=\displaystyle\frac{1}{2}m_{\phi}$). If $m_{\phi} > 2 EeV$, these UHE neutrinos can be detected by the fluorescence detector. The JEM-EUSO telescope is a kind of fluorescence detectors and can be used to detect UHE neutrinos\cite{ebisuzaki}.
\par
UHE neutrinos reach the Earth and pass through the Earth and air, meanwhile these particles interact with matter of the Earth and air. The secondary particles generated by these UHE neutrinos will develop into a cascade. And the most dominant particles in a cascade are electrons moving through atmosphere. Ultraviolet photons are emitted by electron interaction with air. Some of these photons is from nitrogen fluorescence, in which molecular nitrogen excited by a passing shower emits photons isotropically (about 300 - 400 nm). A much larger fraction of these photons is emitted as Cherenkov photons, which are strongly beamed along the shower axis. JEM-EUSO will detect the light from isotropic nitrogen fluorescence and Cherenkov radiation (see Fig. 1). JEM-EUSO consists of photon collecting optics and focal surface detector with its electronics. The optics is composed of three double-sided curved circular Fresnel lenses with 2.65 m maximal diameter. When the ultraviolet light hits the optics it is focused onto the focal surface which consists of 137 Photo-Detector Modules (PDMs). Each PDM consists of a set of elementary cells containing arrays of Multi-Anode Photo Multiplier Tubes with 64 pixels with a spatial resolution of 0.074$^{\circ}$ each\cite{eusomission,Galina}.
\par
JEM-EUSO is a space science observatory to explore the extreme-energy cosmic rays and upward-going neutrinos in the Universe\cite{eusomission}. It will be installed into the International Space Station (ISS from now on) after 2020. The ISS maintains an orbit with an altitude of H $\sim$ 400 km and circles the Earth in roughly 90 minutes. The JEM-EUSO telescope has a wide field of view (FOV: $\pm30^{\circ}$) and observes extreme energy particles in the two modes (nadir and tilted modes) via fluorescent and Cherenkov photons due to the development of extensive air showers. JEM-EUSO is tilted by an angle of 30 degrees in the tilted mode. JEM-EUSO has a observational area of about $2\times10^5$ km$^2$ and $7\times10^5$ km$^2$ in nadir and tilted modes, respectively. The duty cycle for JEM-EUSO, R, is taken to be 10\%. In the present paper, it is made an assumption that there exists air under an altitude of $H_a$ = 100 km.
\par
The number of UHE neutrinos, N$_{det}$, detected by JEM-EUSO can be obtained by the following function:
\begin{center}
\begin{equation}
N_{det} = R\times T\times (A\Omega)_{eff} \times \Phi_{\nu}
\end{equation}
\end{center}
where T is the lifetime of the JEM-EUSO experiment, $\Phi_\nu=\displaystyle\frac{d\psi_\nu}{dE_{\nu}}$ and $(A\Omega)_{eff}$ = the observational area $\times$ the effective solid angle $\times$ P(E,$D_e$,D). Here $P(E,D_e,D)=exp\left(-\displaystyle\frac{D_e}{L_{earth}}\right)\left[1-exp\left(-\displaystyle\frac{D}{L_{air}}\right)\right]$ is the probability that UHE neutrinos interacts with air after traveling a distance between $D_e$ and $D_e+D$, where D is the effective length in the JEM-EUSO detecting zone in the air, $D_e$ are the distances through the Earth and $L_{earth,air}$ are the UHE neutrino interaction lengths with the Earth and air, respectively.
\par
In what follows, $(A\Omega)_{eff}$ is roughly considered and then the numbers of UHE neutrinos detected by JEM-EUSO are evaluated. Here it is made an assumption that the observational area of JEM-EUSO is regarded as a point in the calculation of the effective solid angle $\Omega$. Under this approximation,
\begin{center}
\begin{equation}
(A\Omega)_{eff} \approx A\int_{\theta_{min}}^{\theta_{max}} P(E,D_e,D)\frac{2\pi {R_e}^2sin\theta}{(D_e+D_a)^2} d\theta .
\end{equation}
\end{center}
where $A=\displaystyle\frac{9}{16}S$ (A is the observational area in the case of detection of neutrinos skimming the Earth and S is the observation area of JEM-EUSO), $R_e$ is the radius of the Earth, $D_a$ are the distances through the atmosphere, $\theta$ is the polar angle for the Earth (see Fig. 1), $\theta_{min}$ is the minimum of $\theta$ and $\theta_{max}$ is the maximum of $\theta$. $D_e=2R_ecos\alpha$ and $D_a=(R_e+H_a)cos\beta-R_ecos\alpha$ ($\alpha,\beta$ see Fig. 1). $D=\displaystyle\frac{(H-H_a)sin30^{\circ}}{sin(\beta-30^{\circ})}$ and $D=\displaystyle\frac{(H-H_a)sin60^{\circ}}{sin(\beta-60^{\circ})}$ in the nadir and tilted modes, respectively.
\par
The background due to astrophysical neutrinos is roughly estimated with a diffuse neutrino flux of $\Phi_{\nu}=0.9^{+0.30}_{-0.27}\times(E_{\nu}/100TeV)\times10^{-18}GeV^{-1} cm^{-2}s^{-1}sr^{-1}$\cite{icecube}, where $\Phi_{\nu}$ represents the per-flavor flux, by the above method.

\section{Results}
The numbers of UHE neutrinos, including from superheavy dark matter and astrophysical sources, detected by JEM-EUSO are evaluated at different energy at different $\theta$, respectively. Since JEM-EUSO can only measure the deposited energy E$_{dep}$ in the air, it is important to determine the inelasticity parameter $y=\displaystyle\frac{E_{dep}}{E_{in}}$ (where $E_{in}$ is the incoming particle energy). y for neutrino is about 0.8 at O(10EeV)\cite{gandhi}. If the energy threshold for JEM-EUSO is taken into account (about 20 EeV\cite{threshold}), the energy threshold for neutrino is taken to be 30 EeV at JEM-EUSO in the present paper. Fig. 2 shows the normalized $\theta$ distribution of the neutrino event rates at 30 EeV and 1 ZeV, respectively. From Fig. 2, we can see that neutrinos detected by JEM-EUSO concentrate at large angle. So $\theta_{min}$ is taken to be $160^{\circ}$. $\theta$ reaches its maximum and $\theta_{max} = 169.9^{\circ}$ when a track of a neutrino is tangent to the Earth.
\par
Constrained by the lifetime of ISS, JEM-EUSO has an operation time of five years. The numbers of UHE neutrino, from superheavy dark matter, detected by JEM-EUSO are evaluated at different energy (1 EeV < E < 1 ZeV) in 5 years in the nadir mode when $\tau_\phi=10^{27}$s, respectively (see Fig.3). If the energy threshold for neutrino is taken to be 30 EeV at JEM-EUSO, the numbers of these neutrinos can reach about 1.4$\times10^3$ and 40 at at the energies with 30 EeV and 1 ZeV in 5 years, respectively. The numbers of UHE neutrinos, from superheavy dark matter, detected by JEM-EUSO are evaluated at different energy (1 EeV < E < 1 ZeV) in 5 years in the tilted mode when $\tau_\phi=10^{27}$s, respectively (see Fig.4). The numbers of these neutrinos can reach about 1.9$\times10^4$ and 560 at at the energies with 30 EeV and 1 ZeV in 5 years, respectively. From Fig. 3 and 4, we can see the astrophysical neutrino contamination are less than $10^{-5}$ at energies above 30 EeV. So astrophysical neutrinos could be negligible in this detection.
\par
Since $\Phi_{\nu}$ is proportional to $\displaystyle\frac{1}{\tau_{\phi}}$, the above results are actually depended on the lifetime of superheavy dark matter. For example, the neutrino event rate for JEM-EUSO is $\sim$1.9$\times10^3$ and $\sim$56 events/five years at 30 EeV and 1 ZeV in the tilted mode when $\tau_{\phi}=10^{28}$s, respectively. The neutrino event rate for JEM-EUSO is $\sim$190 and $\sim$6 events/five years at 30EeV and 1 ZeV in the tilted mode when $\tau_{\phi}=10^{29}$s, respectively. The neutrino event rates for JEM-EUSO are $\sim$19 and $\sim$1 events/five years at 30EeV and 500 EeV in the tilted mode when $\tau_{\phi}=10^{30}$s, respectively. The neutrino event rate for JEM-EUSO is $\sim$2 and $\sim$1 events/five years at 30EeV and 50 EeV in the tilted mode when $\tau_{\phi}=10^{31}$s, respectively.
\section{Conclusion}
According to the results described above, it is possible that UHE neutrinos from the decay of superheavy dark matter could be detected at JEM-EUSO under the assumption that superheavy dark matter only decay to neutrinos. It is made an approximation that the observational area is regards as a point in the calculation of the solid angle. This produces some deviations for the event rates of UHE neutrinos, but they can not have an effect on the conclusion that UHE neutrino due to the decay of superheavy dark matter could be detected by JEM-EUSO in 5 years when $O(10^{27})s < \tau_{\phi}< O(10^{31})s$. And it is also possible that these UHE neutrinos are directly probed by the detectors based on ground, such as the Pierre Auger observatory (its observation area $\sim$ $10^3$ km$^3$)\cite{auger}, when $O(10^{27})s < \tau_{\phi}< O(10^{29})s$. The UHE neutrino event rates (due to superheavy dark matter) for the Pierre Auger observatory are $\sim$30 and $\sim$1 events/ten years at 30EeV and 1 ZeV when $\tau_{\phi}=10^{27}$s, respectively. If so, it is confirmed that there may exists superheavy dark matter in the Universe.
\section{Acknowledgements}
This work was supported by the National Natural Science Foundation
of China (NSFC) under the contract No. 11235006, the Science Fund of
Fujian University of Technology under the contract No. GY-Z14061 and the Natural Science Foundation of
Fujian Province in China under the contract No. 2015J01577.
\par

\newpage

\begin{figure}
 \centering
 \includegraphics[width=0.9\textwidth]{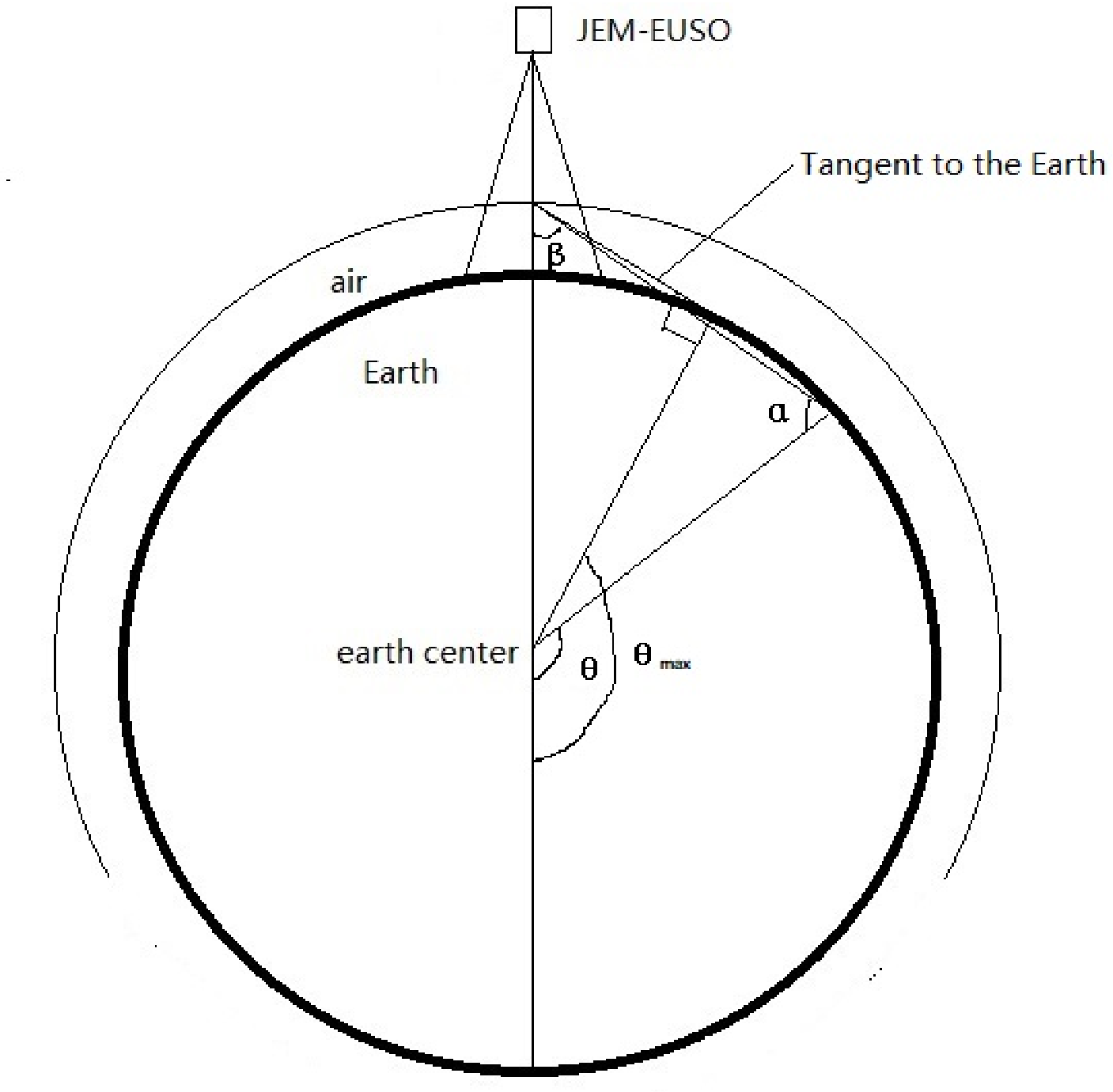}
 \caption{UHE neutrinos skimming the Earth can be detected by JEM-EUSO, via fluorescent and Cherenkov photons due to the development of extensive air showers. $\theta$ is the polar angle for the Earth.}
 \label{fig:figure}
\end{figure}

\begin{figure}
 \centering
 \includegraphics[width=0.9\textwidth]{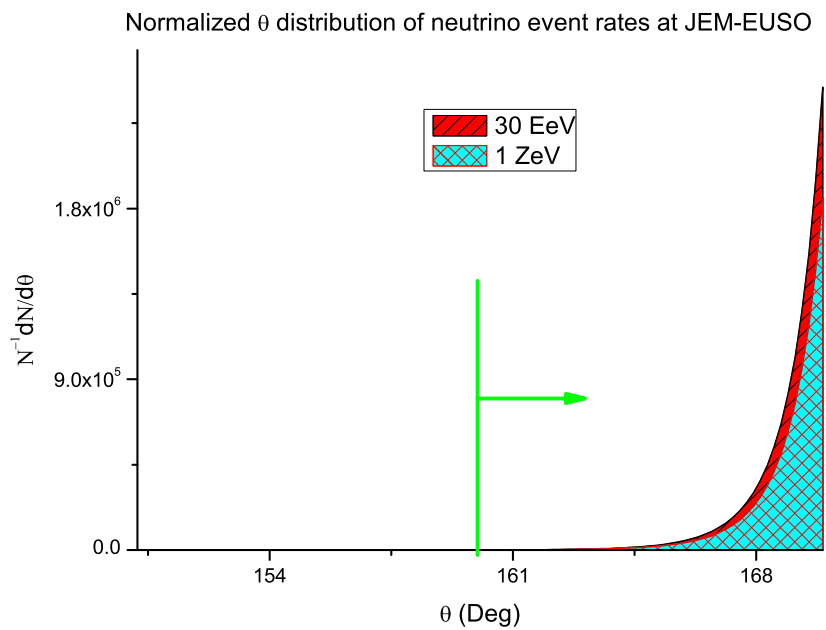}
 \caption{The normalized $\theta$ distribution of the neutrino event rates at 100 EeV and 1 ZeV}
 \label{fig:theta_dis}
\end{figure}

\begin{figure}
 \centering
 \includegraphics[width=0.9\textwidth]{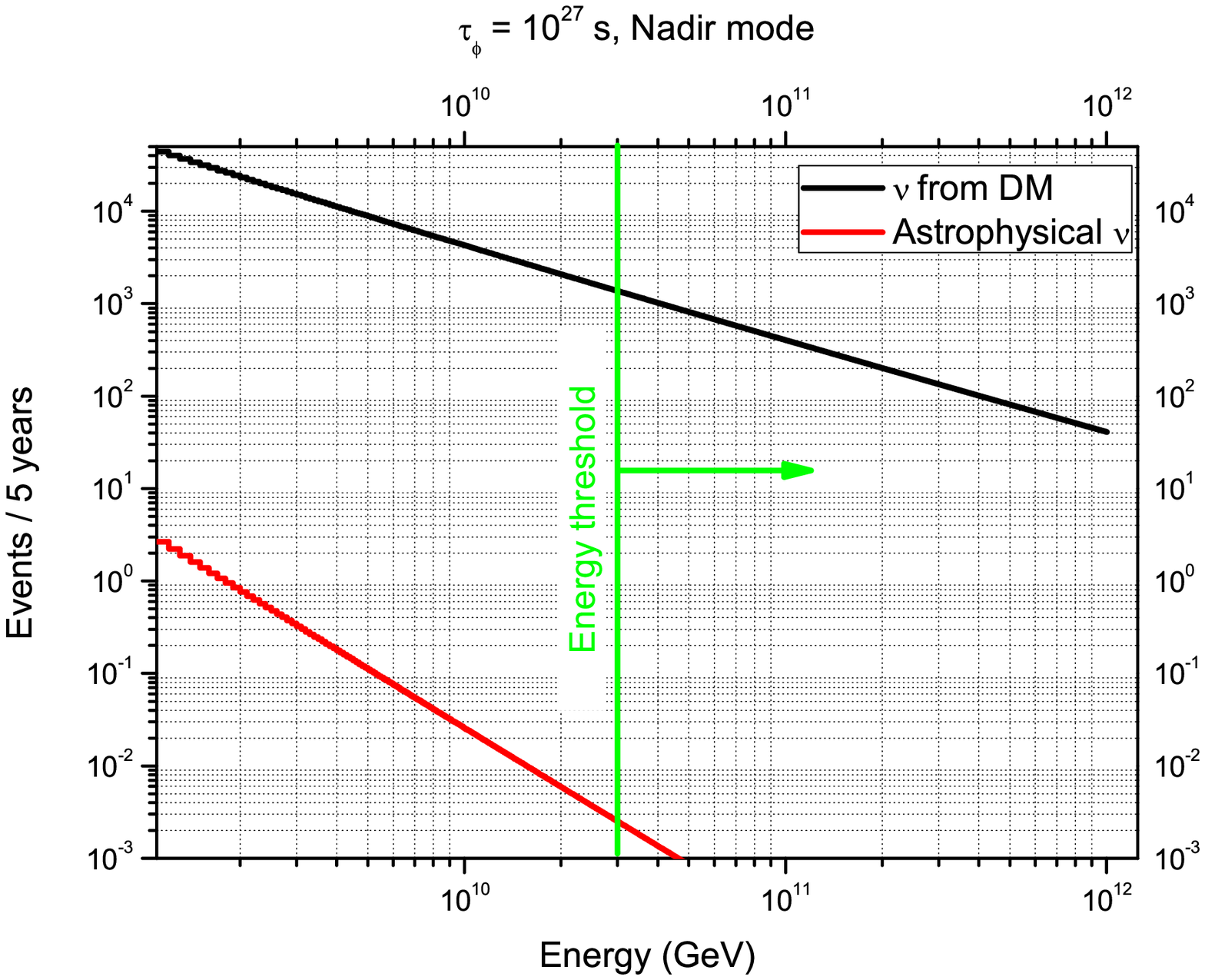}
 \caption{The numbers of UHE neutrino, from superheavy dark matter, detected by JEM-EUSO are evaluated at different energy in 5 years in the nadir mode when $\tau_\phi=10^{27}$s}
 \label{fig:27_nadir}
\end{figure}

\begin{figure}
 \centering
 \includegraphics[width=0.9\textwidth]{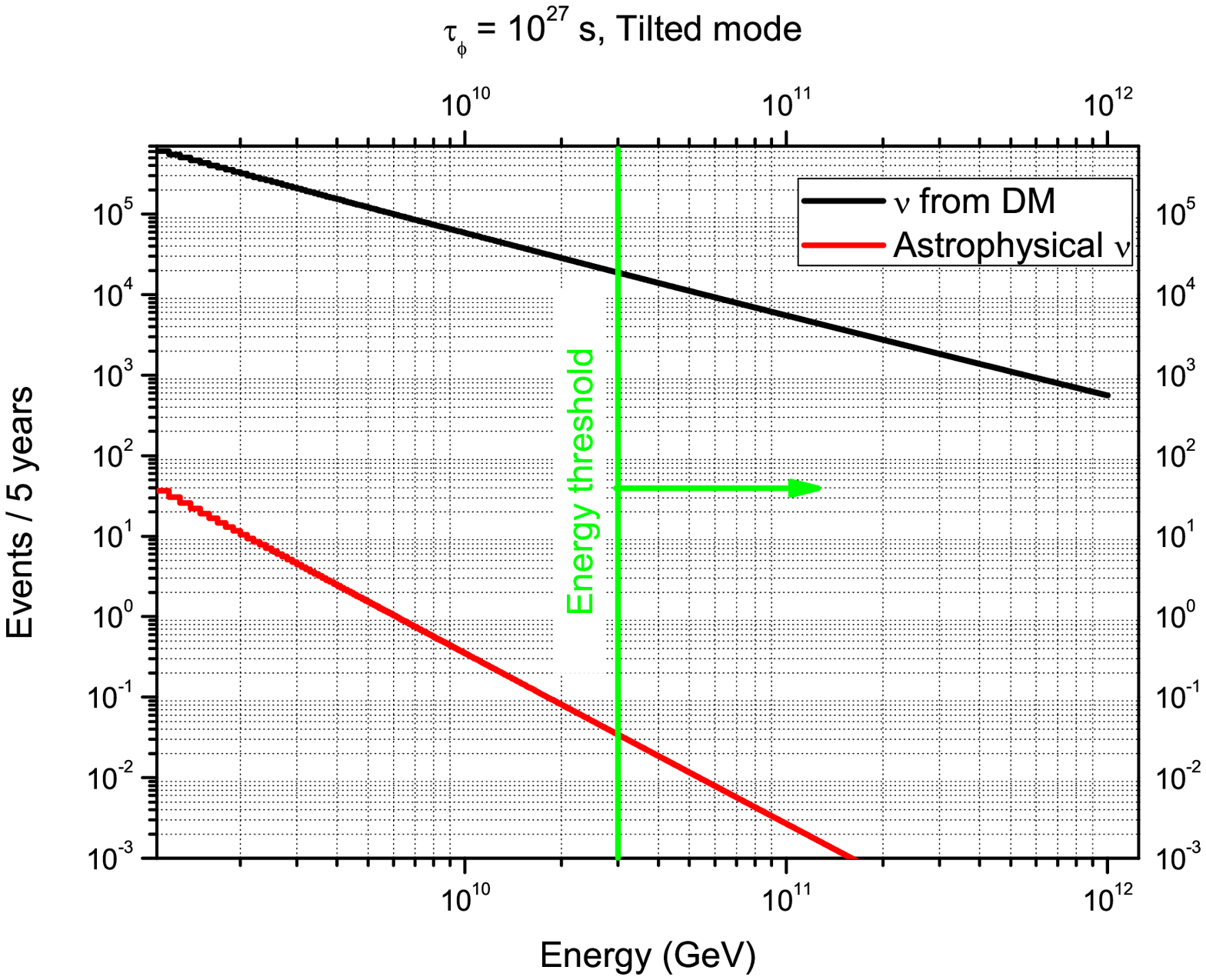}
 \caption{The numbers of UHE neutrino, from superheavy dark matter, detected by JEM-EUSO are evaluated at different energy in 5 years in the tilted mode when $\tau_\phi=10^{27}$s}
 \label{fig:27_tilt}
\end{figure}

\end{document}